\documentclass[iop,apjl,tighten]{emulateapj}
\usepackage{apjfonts} 
\usepackage{rotating}
\usepackage{array}
\usepackage{afterpage}
\usepackage{float}
\usepackage{morefloats}
\usepackage{graphicx}
\usepackage{amsmath}
\usepackage{color}
\usepackage[breaklinks,colorlinks,citecolor=blue,linkcolor=magenta]{hyperref}

\newcommand{\OIII}{\mbox{O\,\textsc{iii}}}

\newcommand{\MgI}{\mbox{Mg\,\textsc{i}}}
\newcommand{\kms}{km s$^{-1}$}

\newcommand{\degree}{$^\circ$}

\slugcomment{ApJ accepted}

\shorttitle{Independence of neutral and ionized gas outflows in low-\lowercase{z} galaxies}
\shortauthors{Bae \& Woo}

\begin{document}

\title{The independence of neutral and ionized gas outflows in low-\lowercase{z} galaxies}

\author{Hyun-Jin Bae$^{1}$}
\author{Jong-Hak Woo$^{1,}$\altaffilmark{2}}

\affil{$^{1}$Astronomy Program, Department of Physics and Astronomy, Seoul National University, Seoul 08826, Republic of Korea; woo@astro.snu.ac.kr}
\altaffiltext{2}{Author to whom any correspondence should be addressed}

\begin{abstract}
Using a large sample of emission line galaxies selected from the Sloan Digital Sky Survey, we investigate the kinematics of the neutral gas in the interstellar medium (ISM) based on the \mbox{Na\,\textsc{i}}$\lambda\lambda$5890,5896 (Na D) doublet absorption line. By removing the Na D contribution from stellar atmospheres, we isolate the line profile of the Na D excess, which represents the neutral gas in the ISM. 
The kinematics traced by the Na D excess show high velocity and velocity dispersion for a fraction of galaxies, indicating the presence of neutral gas outflows. We find that the kinematics measured from the Na D excess are similar between AGNs and star-forming galaxies. Moreover, by comparing the kinematics traced by the Na D excess and those by the [\OIII]$\lambda$5007 line taken from \citet{Woo+17}, which traces ionized outflows driven by AGNs, we find no correlation between them. These results demonstrate that the neutral gas in the ISM traced by the Na D excess and the ionized gas traced by [\OIII] are kinematically independent, and AGN has no impact on the neutral gas outflows. In contrast to [\OIII], we find that the measured line-of-sight velocity shift and velocity dispersion of the Na D excess increase for more face-on galaxies due to the projection effect, supporting that Na D outflows are radially driven (i.e., perpendicular to the major axis of galaxies), presumably due to star formation. 
\end{abstract}


\section{Introduction}
\label{intro}
Gas outflows are frequently observed in star-forming galaxies (SFGs) and active galactic nuclei (AGNs) \citep[e.g.,][]{2010AJ....140..445C,2012ApJ...760..127M,2014MNRAS.441.3306H,2014ApJ...795...30B,2016ApJ...817..108W,2016A&A...588A..41C}. Such outflows transfer vast amount of energy and momentum to the interstellar medium (ISM), plausibly affecting the evolution of the host galaxies \citep[see][for a review]{2005ARA&A..43..769V}. Particularly, AGN-driven outflows may play a crucial role in quenching star formation in massive galaxies as theoretical studies suggest \citep[see][for a review]{2015ARA&A..53..115K}, yet there is a lack of direct observational evidences for the star-formation quenching in AGN-host galaxies \citep[e.g.,][]{2013A&A...549A..43H,VillarMartin:2016ena,Woo+17,Bae:2017js}.

Outflows have been detected in different gas phases, e.g., neutral gas \citep[e.g.,][]{2005ApJS..160...87R,2010AJ....140..445C,le11}, ionized gas \citep[e.g.,][]{2012ApJ...746...86G,2016ApJ...817..108W}, or hot plasma \citep[e.g.,][]{McNamara:2005jm,Aharonian:2016jq}. By examining the kinematics of the [\OIII]$\lambda$5007 line, for example, \citet{2016ApJ...817..108W} showed that the ionized gas outflows are prevalent in AGNs at z$<$0.3,
but not in SFGs \citep[see also][]{2014ApJ...795...30B,Woo+17}. To understand how AGN outflows affect star formation in host galaxies, it is of importance to investigate the cold, neutral gas outflows, which are more closely related to star formation.

The \mbox{Na\,\textsc{i}}$\lambda\lambda$5890,5896 (Na D) doublet is one of the tracers for measuring the properties of neutral gas \citep[e.g.,][]{2005ApJS..160...87R}. Since the Na D line profile is contributed by the neutral gas in the ISM as well as stars, there were various attempts to separate the ISM component in the previous studies. For example, \citet{Heckman:2000du} compared the observed Na D line profiles of 32 far-IR-bright galaxies with those of late-type stars (e.g., K giants) in order to identify galaxies with the interstellar-dominated Na D line. Based on the fact that Na and Mg have similar first ionization potentials, \citet{2005ApJS..160...87R} estimated the stellar contribution in the Na D line profile by utilizing the \MgI\ absorption line profile for their sample of IR-luminous starburst galaxies. As a more sophisticated approach to investigate the outflows in Na D, \citet{2010AJ....140..445C} used two components, respectively representing quiescent disk gas and outflowing gas, by modeling the Na D line profile with physical parameters, velocity and velocity dispersion as well as optical depth and covering factor, using stacked spectra of SFGs. Recently, \citet{Sarzi:2015hf} investigated whether gas outflows in the Na D excess is due to AGN, by exploiting $\sim$450 radio galaxies. They concluded that Na D outflows are driven by star formation, because Na D outflows are detected only among SFGs. Since \citet{Sarzi:2015hf} selected mostly massive early-type galaxies ($\sim$90\%), it is alternatively suggested that the Na D excess may be due to the template mismatch, and that stellar population models with Na-enhanced metallicities may reproduce the observed Na D profile \citep{Jeong:2013kw,Park:2015il}. 

To overcome the limitations in the previous studies and to investigate whether the neutral gas outflows in the ISM are driven by AGN, we investigate the gas outflows traced by the Na D excess using a large and unbiased sample of emission line galaxies selected from the Sloan Digital Sky Survey (SDSS).
In this paper, we present the Na D kinematics for a large sample of SFGs and AGNs, reporting that the neutral gas kinematics are independent of the ionized gas kinematics. In Section \ref{sample}, we describe our sample selection and fitting procedures for the Na D line. In Section \ref{result}, we present the kinematics of Na D excess and compare with [\OIII] kinematics. Finally, we conclude our results in Section \ref{conclusion}.

\begin{figure}
\centering
\includegraphics[width=0.48\textwidth]{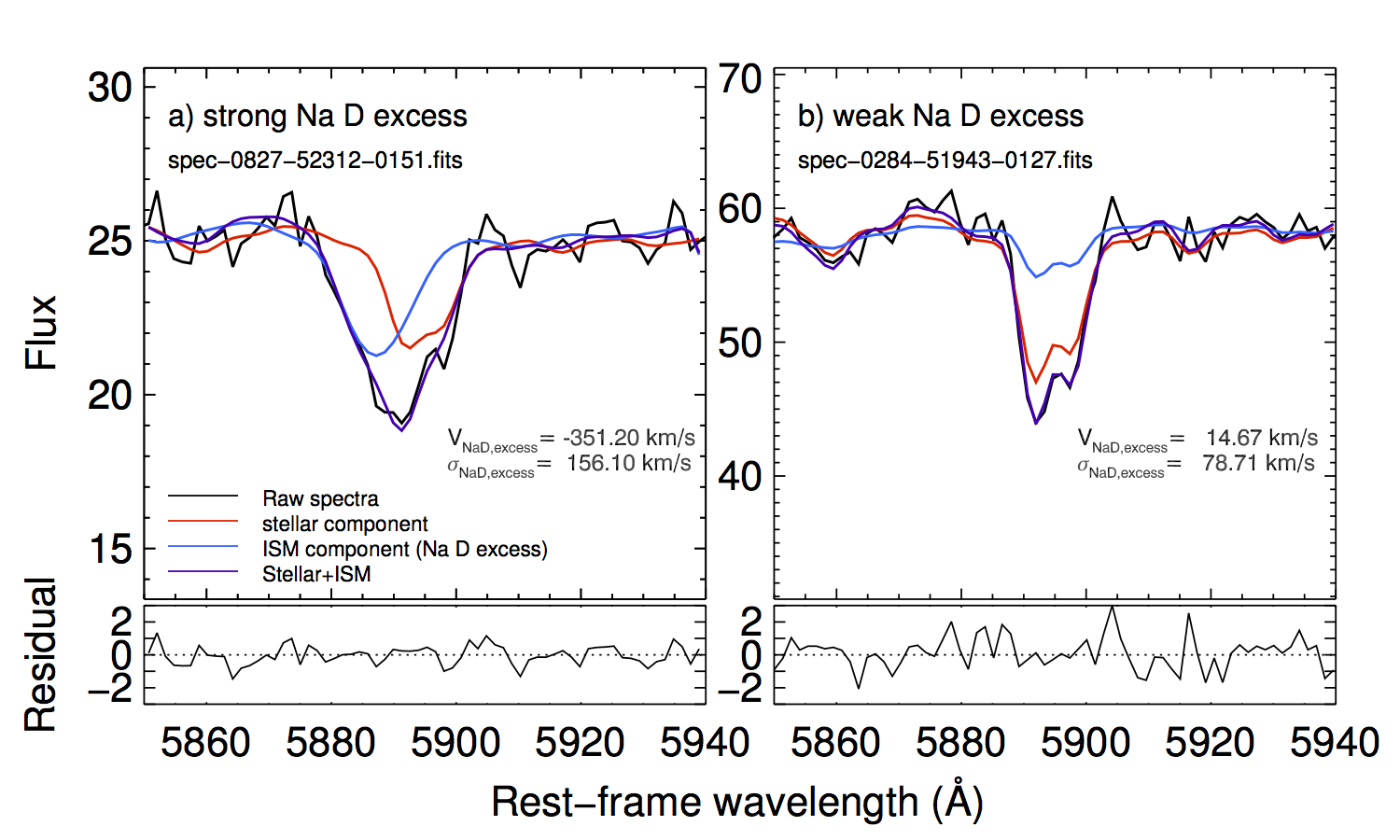}
\caption{Examples of the Na D line fitting using SSP models, for an object with a strong Na D excess (left) and a weak Na D excess (right). The SDSS spectrum (black), the best-fit stellar continuum model (red), the best-fit Na D excess model (blue), and the sum of the best-fit models (purple) are presented in the upper panel while the residual is given in the bottom panel. }
\label{fig:fitting}
\end{figure}

\begin{figure}
\centering
\includegraphics[width=0.48\textwidth]{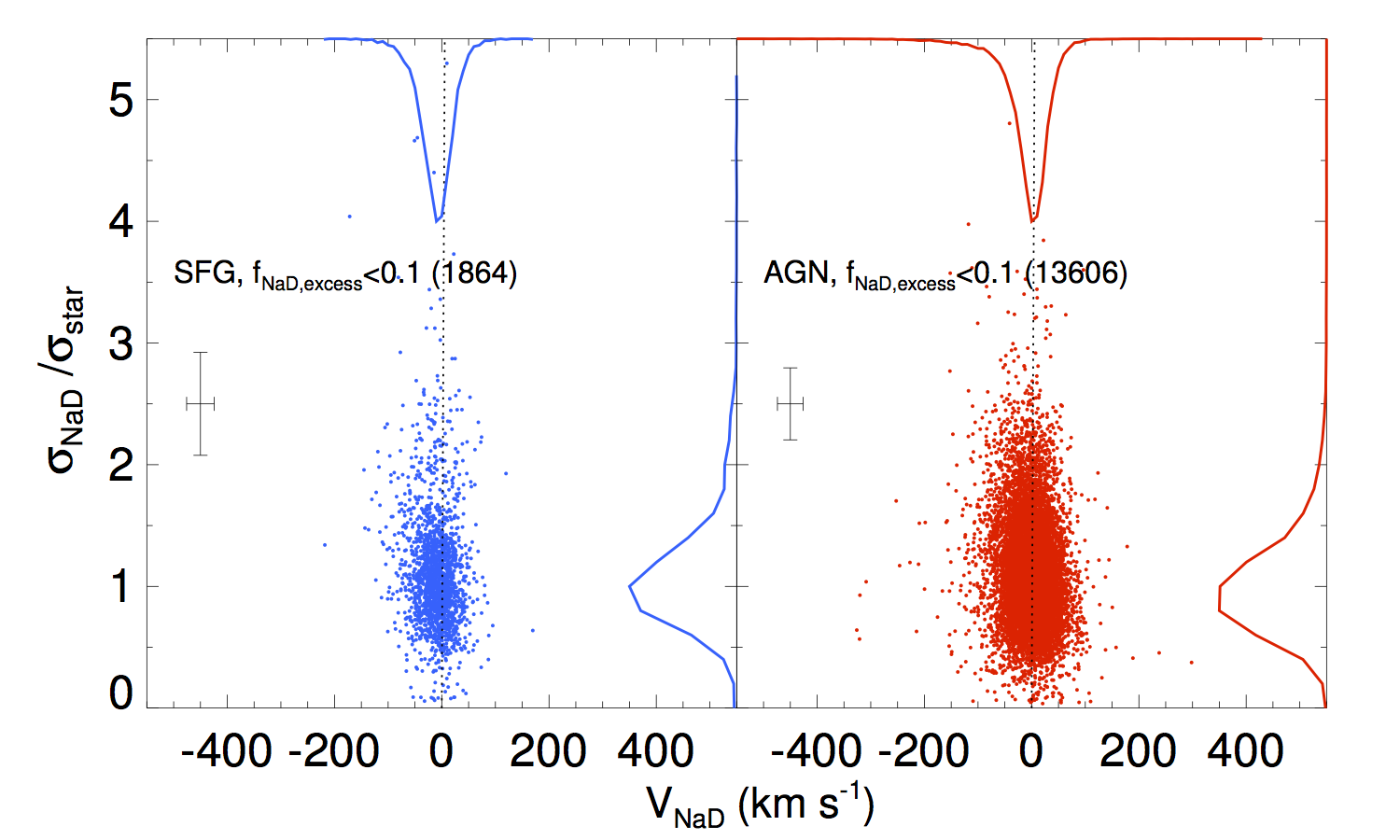}
\includegraphics[width=0.48\textwidth]{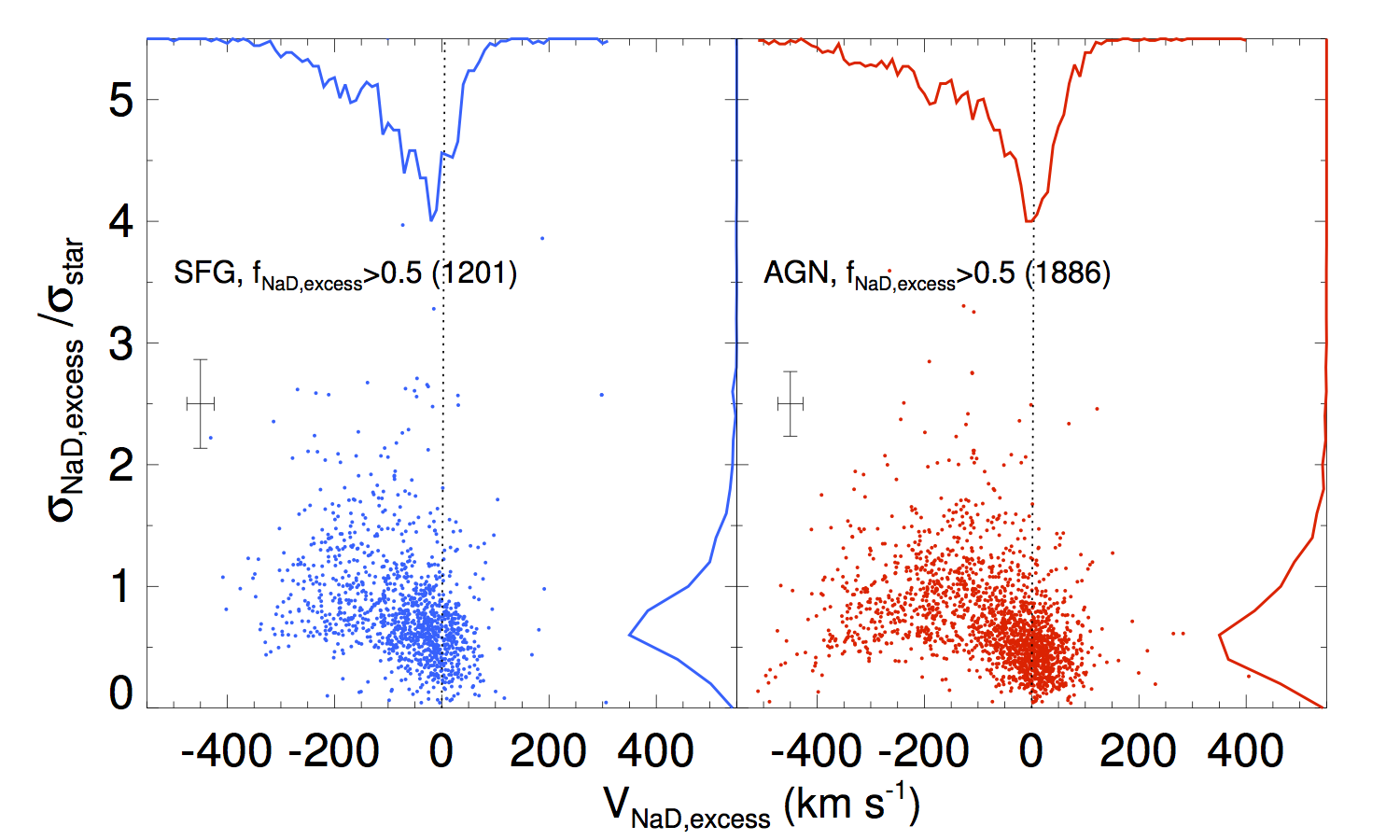}
\caption{Distributions of Na D velocity and normalized velocity dispersion for SFGs (left) and AGNs (right), respectively, with no Na D excess (i.e., $f_{\text{NaD,excess}} < 0.1$, top) and with the Na D excess (i.e., $f_{\text{NaD,excess}} > 0.5$, bottom).  
Normalized histograms are given in each panel. A crosshair in each panel represents a typical 1$\sigma$ uncertainty.}
\label{fig:vvd_nad}
\end{figure}

\section{Sample \& Analysis}
\label{sample}
We utilized a large sample of $\sim$113,000 galaxies with well defined emission lines from \citet{Woo+17} \citep[see also][]{2014ApJ...795...30B,2016ApJ...817..108W, Kang+17}, which is
selected from the SDSS seventh data release (DR7) \citep{2009ApJS..182..543A}.
The sample includes $\sim$43,000 AGNs and $\sim$69,000 SFGs, which are classified based on the emission-line flux ratio diagnostics using [\OIII], H$\beta$, [NII], and H$\alpha$ lines \citep[e.g.,][]{2003MNRAS.346.1055K,2006MNRAS.372..961K}. 
\citet{Woo+17} measured the ionized gas kinematics traced by velocity and velocity dispersion of [\OIII] by modeling the [\OIII] line profile with a single or double Gaussian model. The first and second moments of the best-fit line profile were used to calculated the flux-weighted velocity shifts ($V_{\text{[O III]}}$) with respect to the systemic velocity (i.e., based on stellar absorption lines) and the velocity dispersion ($\sigma_{\text{[O III]}}$), respectively. Then, the properties of ionized gas outflows and their connection to star formation were investigated in detail \citep[see also][]{2014ApJ...795...30B,2016ApJ...817..108W}. 

Using the same sample, we now investigate the neutral gas kinematics by measuring the velocity and velocity dispersion from the Na D excess and directly compare them with the ionized gas kinematics.
The Na D line often shows a larger line strength than other stellar lines due to the contribution from ISM. We will refer this component as the Na D excess. Thus, in order to measure the kinematics of neutral gas in the ISM, we need to separate the ISM component from the stellar component in the Na D line profile. For separating the Na D excess component, we fist fitted the stellar continuum over the spectral range 3700$-$7600\AA, which contains multiple strong stellar absorption lines, using the penalized pixel fitting code \citep[pPXF,][]{2004PASP..116..138C}. We utilized a combination of simple stellar population (SSP) models \citep[MILES,][]{2006MNRAS.371..703S}, adopting solar metallicity and age spanning from 60 Myr to 12.6 Gyr. While we performed the fitting process with/without masking the Na D line, we find no significant difference since many other strong absorption lines are dominant in the fitting process. 

Second, after subtracting the best-fit stellar population model, we fitted the Na D excess at a fixed wavelength range 5850--5940\AA\ using the pPXF. Note that the Na D excess represents the neutral gas in the ISM, which is presumably composed of non-outflowing quiescent gas in the disk as well as outflowing gas. We did not attempt to separate these two components since they are degenerate. Instead, we used the total line profile of the Na D excess for measuring the velocity and velocity dispersion to constrain the kinematics of the neutral gas in the ISM. To avoid any contamination, we masked the \mbox{He\,\textsc{i}}$\lambda$5876 line  in the fitting process.
Based on the best-fit model for the NaD excess, we measured the velocity shift  ($V_{\text{NaD,excess}}$) with respect to the systemic velocity, which is measured from the stellar continuum in the previous step, and velocity dispersion ($\sigma_{\text{NaD,excess}}$) of the Na D excess. For comparison, we also used the observed profile of the Na D line, without subtracting the stellar component and obtained kinematic measurements, $V_{\text{NaD}}$ and $\sigma_{\text{NaD}}$. While our fitting method is different from that of the previous studies \citep[e.g.,][]{2005ApJS..160...87R,2010AJ....140..445C,le11,Sarzi:2015hf}, our method is straightforward to use for comparing with the kinematics of [\OIII] and provides relatively good fitting results (see Figure \ref{fig:fitting}). Since we focus on the Na D kinematics, a different choice of SSP models (e.g., initial mass function, metallicity, or age) does not affect the results. 

We measured the equivalent width (EW) of the Na D absorption line for both the excess component and the total profile, and obtained the uncertainty of EWs using Monte Carlo simulations based on flux errors. The pseudo-continuum around Na D was defined using the mean flux in the wavelength ranges, 5850--5870\AA\ and 5920--5940\AA. Based on the Monte Carlo simulation results using a small random subsample, we found that the uncertainties of Na D velocity shift and velocity dispersion strongly depend on the Na D EW. Since it is extremely time-consuming to perform Monte Carlo simulations for all galaxies, we alternatively obtained an empirical functional form for assigning the uncertainty of the Na D velocity shift and velocity dispersion as a function of the Na D EW. Thus, we adopted the representative uncertainties of the Na D velocity and velocity dispersion, using the measured Na D EW of each galaxy. 

To avoid unreliable measurements, we focus on a sample of galaxies that have a strong Na D absorption line, i.e., the EW of total Na D is larger than three times of its uncertainty. We further rejected 12 AGNs and 3 SFGs with a complex Na D profile, which was difficult to fit using our method. Note that the removal of these 15 galaxies does not affect our results. As a final sample, we used 27,013 AGNs ($\sim$62\% of total AGN sample) and 6,783 SFGs ($\sim$10\% of total SFG sample) in this study. Among them, we selected galaxies with the Na D excess if the fraction of the Na D excess is more than 50\%, i.e., $f_{\text{NaD,excess}} > 0.5$, where the Na D excess fraction, $f_{\text{NaD,excess}}$ is defined by the ratio between EW of Na D excess and EW of the total Na D line (i.e., EW (Na D excess) / EW (Na D total)). The number of the Na D-excess galaxies with $f_{\text{NaD,excess}} > 0.5$ is 1,201 among SFGs ($\sim$18\% of the selected SFGs) and 1,886 among AGNs ($\sim$7\% of the selected AGNs). Note that even if we select galaxies with
a lower Na D-excess fraction, e.g., $f_{\text{NaD,excess}} > 0.3$, we obtained qualitatively the same results. In addition to the Na D-excess galaxies, we also selected galaxies with little Na D excess, i.e., $f_{\text{NaD,excess}} < 0.1$. These galaxies are composed of 1,864 SFGs, $\sim$27\% of the selected SFGs, and 13,606 AGNs, $\sim$50\% of the selected AGNs. This fraction indicates that approximately $\sim$50\% of AGNs show no significant signature of the ISM component in the Na D line profile. We will utilize the kinematics measured from the Na D excess to examine the ISM gas kinematics in the next section.

\section{Result}
\label{result}
\subsection{Na D Kinematics}
We compare the Na D kinematics between galaxies with and without the Na D excess in Figure \ref{fig:vvd_nad}. 
For the galaxies without the Na D excess (top panel), the velocity shift of the total Na D line ($V_{\text{NaD}}$) is negligible with the mean value close to 0 \kms\ for both SFGs ($-$11$\pm$32 \kms)\footnote{Note that hereafter we quote the mean and the 1-$\sigma$ dispersion of the distribution} and AGNs (0$\pm$32 \kms), confirming that there is no velocity difference between Na D and other stellar lines. Regardless of AGN activity, the velocity measured from multiple stellar lines is consistent with that measured from the Na D line only. Note that since the typical uncertainty of $V_{\text{NaD}}$ measurement is $\sim$23 \kms\ based on our Monte Carlo simulations, the distribution of $V_{\text{NaD}}$ has a small width as shown in Figure \ref{fig:vvd_nad}.
In the case of velocity dispersion, the ratio between Na D ($\sigma_{\text{NaD}}$) and the stellar lines ($\sigma_{\text{star}}$) is close to unity for both SFGs (1.12$\pm$0.49) and AGNs (1.09$\pm$0.40), indicating that the Na D absorption line is
mainly contributed from stars. As expected we find no difference of Na D kinematics between AGNs and SFGs
since AGN activity has no impact on the stellar kinematics.

In contrast, we find strong signatures of the neutral gas outflows in the Na D excess galaxies (bottom panel). The distribution of the velocity shift measured from the Na D excess component ($V_{\text{NaD,excess}}$) has a long tale to negative values, indicating that many of these objects show a blue-shifted Na D excess profile (see Figure \ref{fig:fitting} left). The mean $V_{\text{NaD,excess}}$ is $-$75$\pm$97 \kms\ 
and $-$77$\pm$122 \kms, respectively for SFGs and AGNs, indicating that strong outflows of the neutral gas are present. We perform the Kolmogorov-Smirnov (K-S) test to investigate whether SFGs and AGNs have the same $V_{\text{NaD,excess}}$ distribution. The results show that the two distributions are not same (K-S statistic = 0.108, p < 0.001), presumably due to 
the difference of the host galaxy mass and star formation rate, etc. Nevertheless, the similar mean and dispersion of the two distributions suggest that the origin of the outflows are similar.

In the case of the velocity dispersion of the Na D excess component, the mean ratio between $\sigma_{\text{NaD,excess}}$ and $\sigma_{\text{star}}$ is less than one for both SFGs (0.83$\pm$0.57) and AGNs (0.72$\pm$0.46), suggesting that the non-gravitational kinematic component traced by the Na D excess is weaker than the gravitational component traced by stellar absorption lines.  Although the distributions are not statistically the same (K-S statistic = 0.148, p < 0.001), the two distributions are similar. It is not likely that AGN activity caused the neutral gas outflows in the ISM since SFGs also show similar outflows in the ISM.

These results may indicate that neither AGN nor star-formation activities are a dominant source of neutral gas outflows in the ISM, as similarly reported by previous studies with various samples \citep[e.g.,][]{Colbert+96}. On the other hand, since AGNs with strong [\OIII] outflows tend to have relatively high star formation rate as star-forming main sequence galaxies \citep{Woo+17}, it is also reasonable to assume that the neutral gas outflows are caused by star formation activity both in SFGs and AGNs.

\begin{figure}
\centering
\includegraphics[width=0.48\textwidth]{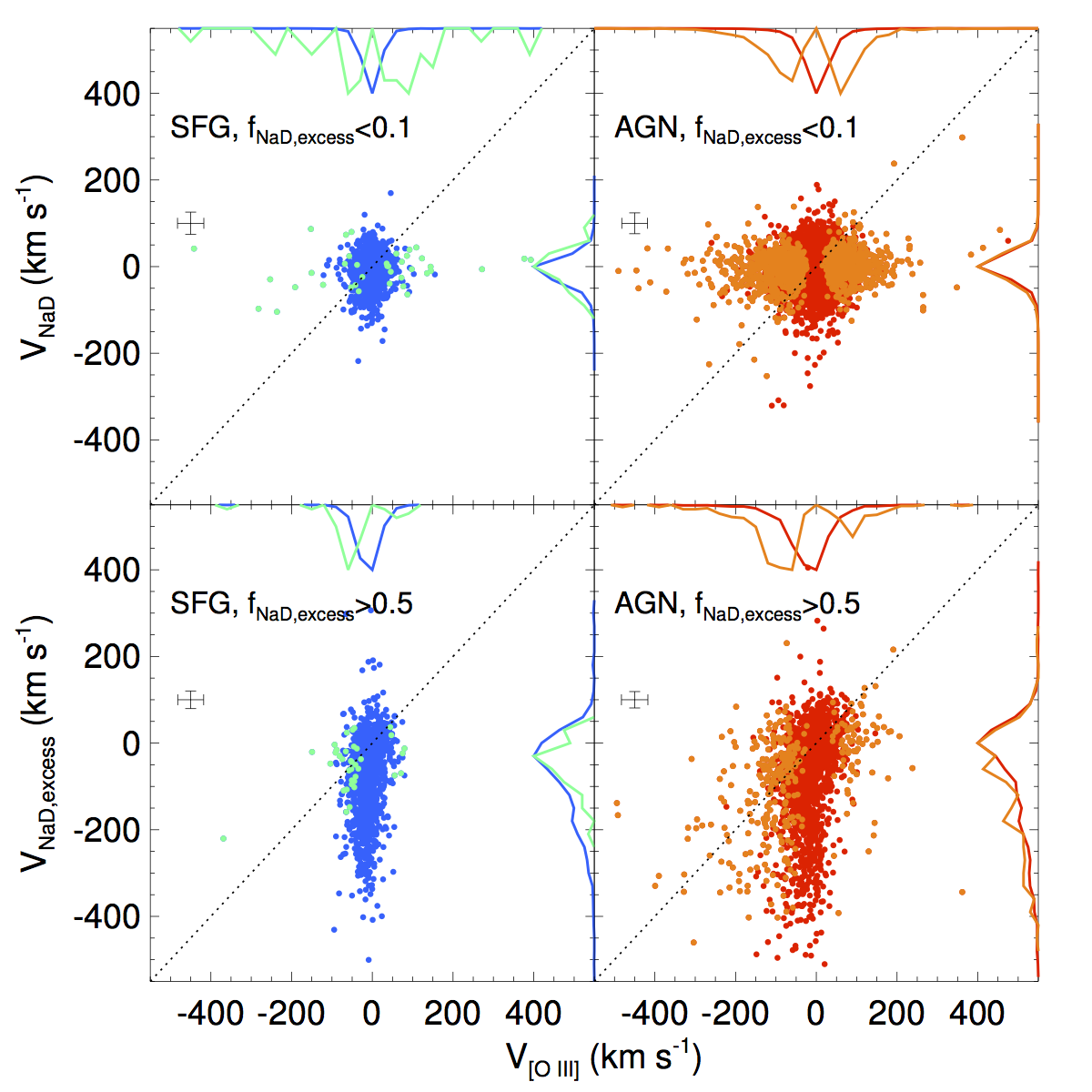}
\caption{Comparison between the Na D velocity and the [\OIII] velocity for SFGs (left) and AGNs (right), respectively, with no Na D excess (i.e., $f_{\text{NaD,excess}} < 0.1$, top) and with the Na D excess (i.e., $f_{\text{NaD,excess}} > 0.5$, bottom).  
We overlaid the sample with robust [\OIII] measurements, i.e., [\OIII] velocity is greater than three times of the uncertainty (orange/cyan dots for AGNs/SFGs, respectively). 
Normalized histograms are given in each panel. A crosshair in each panel represents a typical 1$\sigma$ uncertainty.
}
\label{fig:comp1}
\end{figure}

\begin{figure}
\centering
\includegraphics[width=0.48\textwidth]{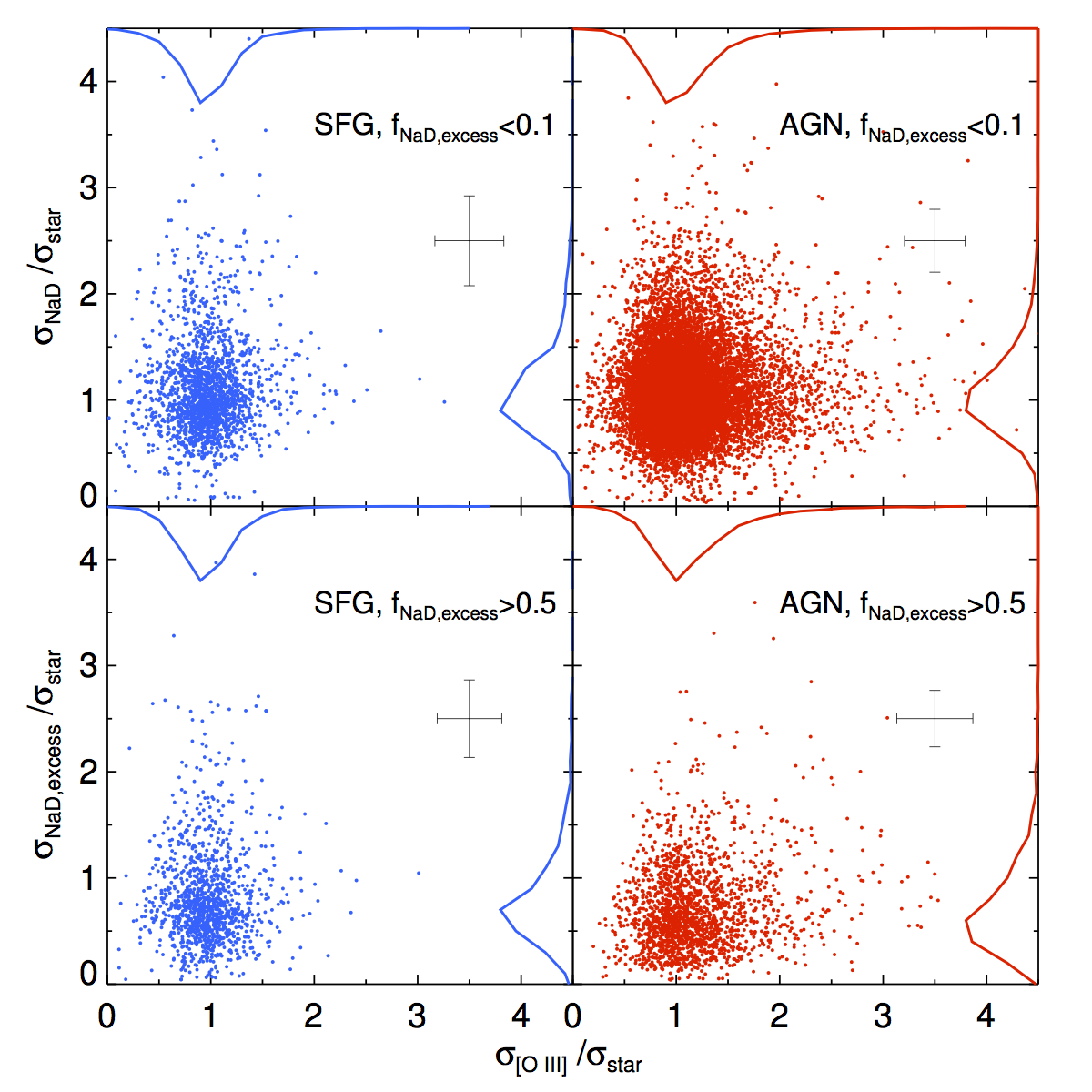}
\caption{Comparison between the Na D velocity dispersion and the [\OIII] velocity dispersion for SFGs (left) and AGNs (right), respectively, with no Na D excess (i.e., $f_{\text{NaD,excess}} < 0.1$, top) and with a strong Na D excess ($f_{\text{NaD,excess}} > 0.5$, bottom).  
Normalized histograms are given in each value. A crosshair in each panel represents a typical 1$\sigma$ uncertainty.
}
\label{fig:comp2}
\end{figure}

\subsection{Comparison to [O III] kinematics}

In this section we directly compare the kinematics of Na D with the kinematics traced by the [\OIII] line, which represent the ionized gas outflows driven by AGNs, by adopting [\OIII] velocity shift and velocity dispersion measurements from \citet{Woo+17}.
If AGNs affect both ionized and neutral gas at the same time, we expect a correlation between the kinematics of the Na D excess and [\OIII]. 
First, we compare the velocity shifts of Na D with that of [\OIII] in Figure \ref{fig:comp1}, finding no correlation between them.
For SFGs without the Na D excess (top-left), the distribution is narrow and negligible for both [\OIII] and Na D (i.e., $V_{\text{[O III]}}$ = $-$2$\pm$33 \kms, and $V_{\text{NaD}}$ = $-$11$\pm$32 \kms, respectively), indicating no outflows in [\OIII] or Na D as expected.
For AGNs without the Na D excess (top-right), the distribution of [\OIII] velocity shift is somewhat broader ($V_{\text{[O III]}}$= 0$\pm$46 \kms) than that of SFGs, due to the presence of AGN driven ionized gas outflows, while the distribution of Na D velocity shift is similar to that of SFGs ($V_{\text{NaD}}$ = 0$\pm$32 \kms), confirming our expectation that there is no AGN effect on the stellar Na D lines.   
Since the majority of galaxies show a weak velocity shift in [\OIII] due to the projection effect since the direction of AGN outflows is expected to be perpendicular to the line-of-sight for type 2 AGNs \citep[see the detailed model prediction based on the Monte Carlo simulation by][]{2016ApJ...828...97B}, we focus on a subsample with the robust  $V_{\text{[O III]}}$ measurements (i.e., $V_{\text{[O III]}}$ $>$ $3\sigma$).
For this subsample (cyan and orange points in Figure 3), we find that $V_{\text{[O III]}}$ is significantly larger in AGNs than in SFGs. This trend is consistent with general population of SFGs and AGNs (see \citet{Woo+17} for the detailed comparison of [\OIII] kinematics between AGNs and SFGs). In contrast, there is no difference in 
the distribution of $V_{\text{NaD}}$ between AGNs and SFGs since stellar kinematics are not affected by AGN activity or star formation. 

Now, we turn to galaxies with the Na D excess, which represents the kinematics of the neutral gas in the ISM (bottom panels in Figure 3). For SFGs with the Na D excess (bottom-left), the distribution of $V_{\text{NaD,excess}}$ is skewed with a long tale to negative values ($V_{\text{NaD,excess}}$= $-$75$\pm$97 \kms), suggesting that some fraction of SFGs have strong outflows. In contrast, $V_{\text{[O III]}}$ shows on average zero velocity shift with a narrow distribution ($V_{\text{[O III]}}$=$-$11$\pm$29 \kms) as expected since there is no AGN. In other words, the neutral outflows traced by the Na D excess can be strong even without AGN activity. 
In the case AGNs with the Na D excess (bottom-right), the distribution of $V_{\text{[O III]}}$ is much broader and blueshifted ($V_{\text{[O III]}}$= $-$25$\pm$70 \kms) than that of SFGs with the Na D excess as expected from the fact that [\OIII] outflows are driven by AGNs. 
In contrast, we find no significant difference of $V_{\text{NaD,excess}}$ between AGNs and SFGs. The distribution
of $V_{\text{NaD,excess}}$ of AGNs is shifted to negative values with a mean of $-$77$\pm$122 \kms, which is similar to that of SFGs with the Na D excess. 
We find no clear correlation between $V_{\text{[O III]}}$ and $V_{\text{NaD,excess}}$ in AGNs, indicating that the neutral gas outflows traced by the Na D excess
and the ionized gas outflows driven by AGN is kinematically independent. 
 
Second, we compare the velocity dispersion of Na D and [\OIII] after normalizing them by stellar velocity dispersion ($\sigma_{\text{star}}$) in Figure \ref{fig:comp2}. We find no clear correlation between Na D and [\OIII] velocity dispersions in any subsample. In the case of AGNs, there is a large range of [\OIII] velocity dispersion for given low velocity dispersion of Na D excess. At the same time, for given low velocity dispersion of [\OIII], there is also a large range of Na D velocity dispersion. These results  indicate that neutral gas outflows manifested by the Na D excess is not directly connected to AGN outflows. 

By comparing SFGs with/without the Na D excess (left panels in Figure 4), we find no significant distribution of the $\sigma_{\text{[O III]}}$. For example, the mean and rms dispersion of the distribution is 0.97$\pm$0.30 and 0.98$\pm$0.31, respectively, for SFGs with/without the Na D excess. In other words, there is no strong ionized gas outflows traced by [\OIII], regardless of the presence of ISM component in the Na D line. In the case of Na D, the distribution of the $\sigma_{\text{Na D,excess}}$ (mean value of 0.83$\pm$0.57) is different from that of the $\sigma_{\text{Na D}}$ (mean value of 1.12$\pm$0.49), reflecting that the Na D excess component represents non-gravitational outflow kinematics. In the case of AGNs, we find no difference of the $\sigma_{\text{[O III]}}$ distribution between AGNs with/without the NaD excess (right panels in Figure 4), suggesting that the kinematics of [\OIII] and the Na D excess have different origins.

When we compare SFGs and AGNs, we find qualitatively the same trend of the distribution of the normalized $\sigma_{\text{Na D,excess}}$ between AGNs and SFGs, while the distribution of $\sigma_{\text{[O III]}}$ is different between SFGs and AGNs as there is a long tail toward the high velocity dispersion. These results indicate that the kinematics of the Na D excess is apparently independent of AGN activity.

\begin{figure}
\centering
\includegraphics[width=0.5\textwidth]{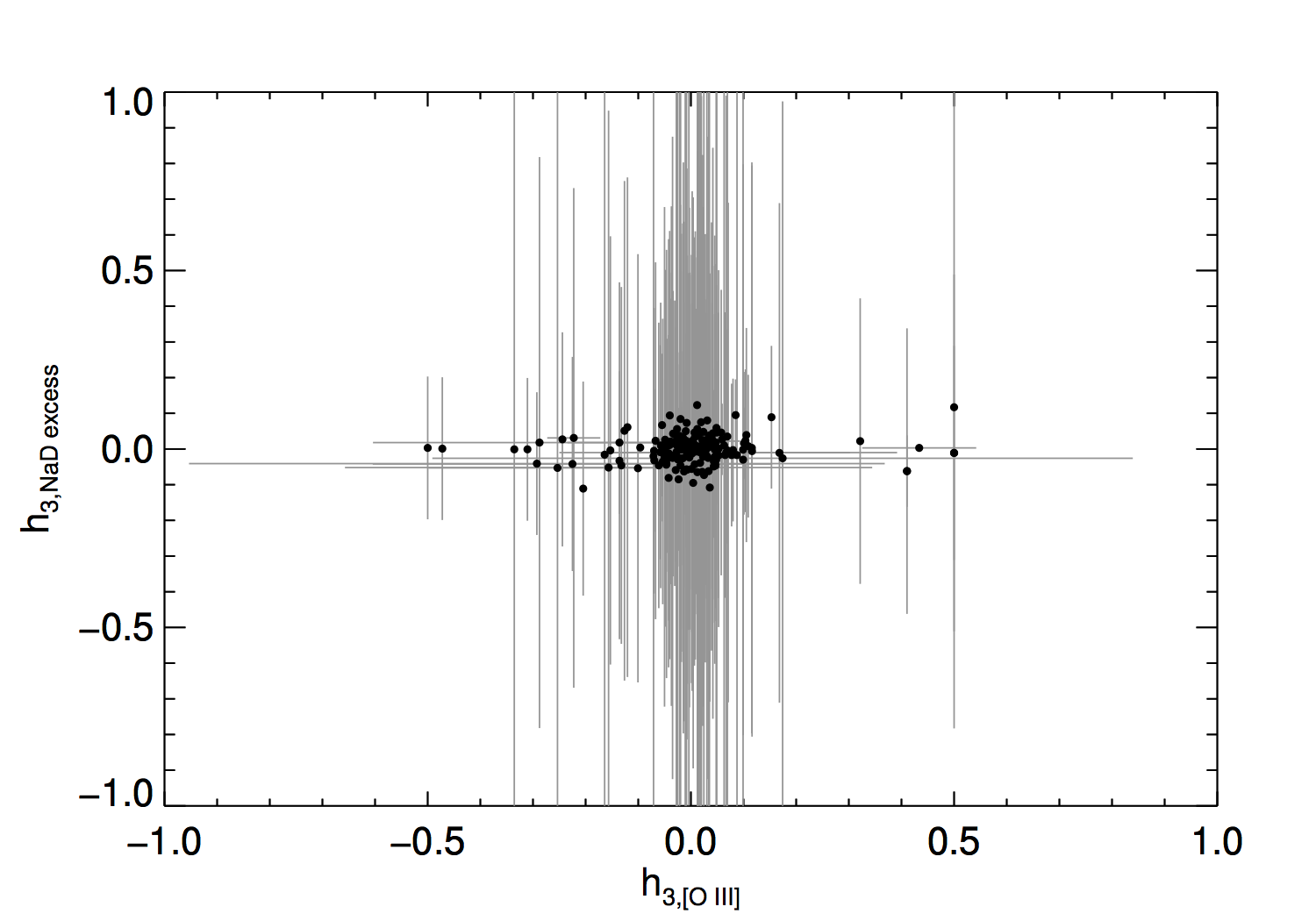}
\includegraphics[width=0.5\textwidth]{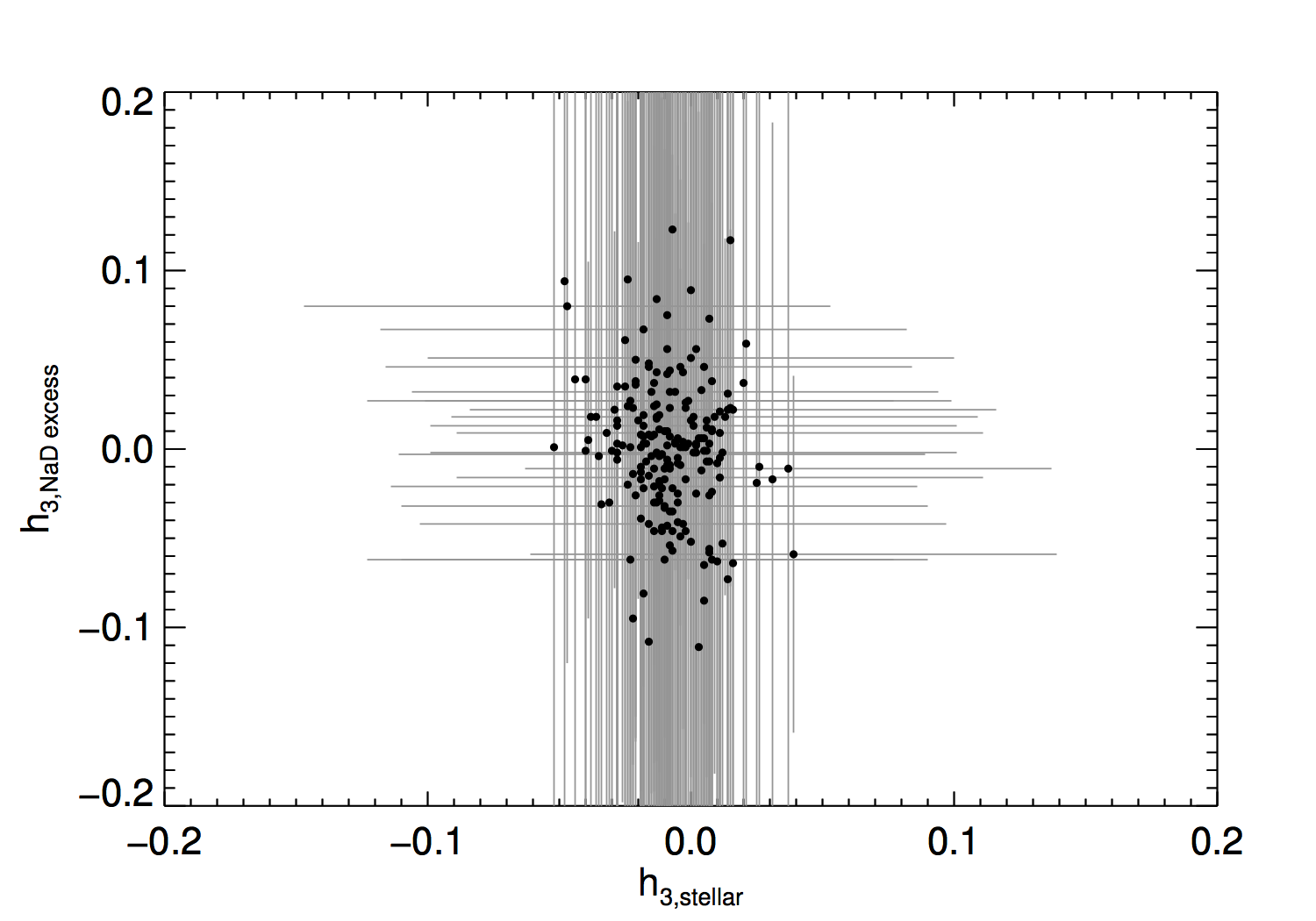}
\caption{Comparisons of the skewness coefficient $h_3$ for AGNs with a high S/N 1) between the Na D excess and the stellar lines (top), 2) between the Na D excess and [\OIII] (bottom). The gray lines represent formal errors of the $h_3$ coefficient for each AGN.  \label{fig:h3comp}}
\end{figure}

\subsection{Comparison of the line skewness}

In this section, we compare the line skewness of the Na D excess and [\OIII]. The line skewness can be measured by fitting the line profile using Gauss-Hermite series as follow:
\begin{eqnarray}
L(y) = {Ae^{-y^2/2}\over \sigma \sqrt{2\pi}}\bigg[1+ \sum_{m=3}^{M}h_m H_m (y) \bigg]
\end{eqnarray}
where $H_3 = y (2y^2 -3) / \sqrt{3}$, $H_4 = [4(y^2-3)y^2 +3] / \sqrt{24}$, and $y \equiv (v-V)/\sigma$. $A$ is the amplitude of line, $V$ is the mean radial velocity, and $\sigma$ is the velocity dispersion of line if we assume the line profile to be a Gaussian \citep{vdm93}. Note that if both $h_3$ and $h_4$ equal to zero, the profile becomes a Gaussian. 

In practice we measured the skewness coefficient ($h_3$) of the Na D excess and the [\OIII] line, using the pPXF and PROFIT \citep{ri10}, respectively.
Since the measurement of skewness is sensitive to the quality of spectra, we selected $\sim$200 AGNs, that have a sufficient signal-to-noise ratio (i.e., S/N $>$ 50) for the skewness analysis, and measured the $h_3$ coefficient in Eq. 1. For this subsample, we measured the $h_3$ of the Na D excess, [\OIII],
and stellar lines, respectively, and compare them.

In the case of Na D excess, the $h_3$ coefficient ranges in $\sim$$\pm$0.1, with a mean, 0.00$\pm$0.04, indicating that the Na D-excess profiles do not show strong skewness. Since the Na D excess represents the ISM gas, which is composed of quiescent and outflowing components, the mean value close to zero indicates that quiescent gas may be dominant. On the other hand, if the outflowing gas is dominant as manifested by a strong blue shift, the line profile is also symmetric with weak skewness. In the case of stars, the $h_3$ coefficient from stellar lines shows a mean value of $-$0.01$\pm$0.02 with a narrow range of $\pm$$\sim$0.05, which is consistent with the expectation from a symmetric line-of-sight velocity distribution of stars in a galactic bulge.  Thus, compared to stars, neutral gas traced by Na D excess shows somewhat larger distribution of skewness, reflecting the non-gravitational outflows. 

In the case of [\OIII], we measure the mean $h_3$ coefficient as 0.00$\pm$0.12, indicating the skewness in the [\OIII] line profile is on average small. This result is consistent with the distribution of [\OIII] velocity shift shown in Figure 3, reflecting that a majority of type 2 AGNs show very small flux-weighted velocity shift \citep{2014ApJ...795...30B,2016ApJ...817..108W, Woo+17}. The trend of skewness of [\OIII] is similar to that of velocity shift since both of them represent the asymmetry of the line profile. 
On the other hand, we find that the $h_3$ range of [\OIII] is much broader than Na D, up to $\pm$$\sim$0.5 (bottom panel in Figure 5),
indicating that there are AGNs with a strongly skewed [\OIII] line, reflecting ionized gas outflows. 

Since the uncertainty of $h_3$ is very large, we focus on AGNs with relatively large $h_3$, i.e., $|h_3|>$0.1, which  is 18\% of the subsample of $\sim$200 AGNs. When we compare the distribution of $h_3$ coefficient between the Na D excess and [\OIII] using these AGNs, we find no correlation between the $h_3$ coefficients of the Na D excess and [\OIII] (bottom panel in Figure \ref{fig:h3comp}).
In other words, while [\OIII] manifests outflows by the skewness of the line profile, the Na D excess does not exhibit outflows as strong as [\OIII].
The results further support the independence of kinematics between neutral and ionized gas outflows in AGNs. 

\subsection{Gas kinematics vs. host galaxy's inclination}
\label{incl}
The independence of the kinematics between the Na D excess and [\OIII] suggests that the neutral gas outflows in the ISM is not related to AGN activity. Instead, star formation-driven outflows may be responsible for the detected kinematics of the Na D excess \citep[e.g.,][]{Sarzi:2015hf}. In this case, the measured velocity shift and velocity dispersion along the line-of-sight may depend on host galaxies' inclination since the direction of star formation-driven outflows is along the rotational axis of host galaxies, while AGN-driven outflows have a random direction with respect to the stellar disk \citep[e.g.,][]{2005ARA&A..43..769V,2014ApJ...795...30B}. In fact, previous studies reported that the strength of ISM outflows correlates with the inclination of the galaxies. For example, using a small of 32 starburst galaxies, \cite{Heckman:2000du} reported that the outflows indicated by the velocity shift of the Na D excess are systematically found in more nearly face-on galaxies than other galaxies, concluding that the outflows are in the direction of the minor axis of the galaxies. Using a larger sample of SDSS SFGs, \cite{2010AJ....140..445C} showed that the line-of-sight outflow velocity is more or less constant when the inclination of the galaxies is more face-on (i.e., $<$ 60 \degree), consistent with the radial direction (perpendicular to the disk) of gas outflows. 

\begin{figure}
\centering
\includegraphics[width=0.48\textwidth]{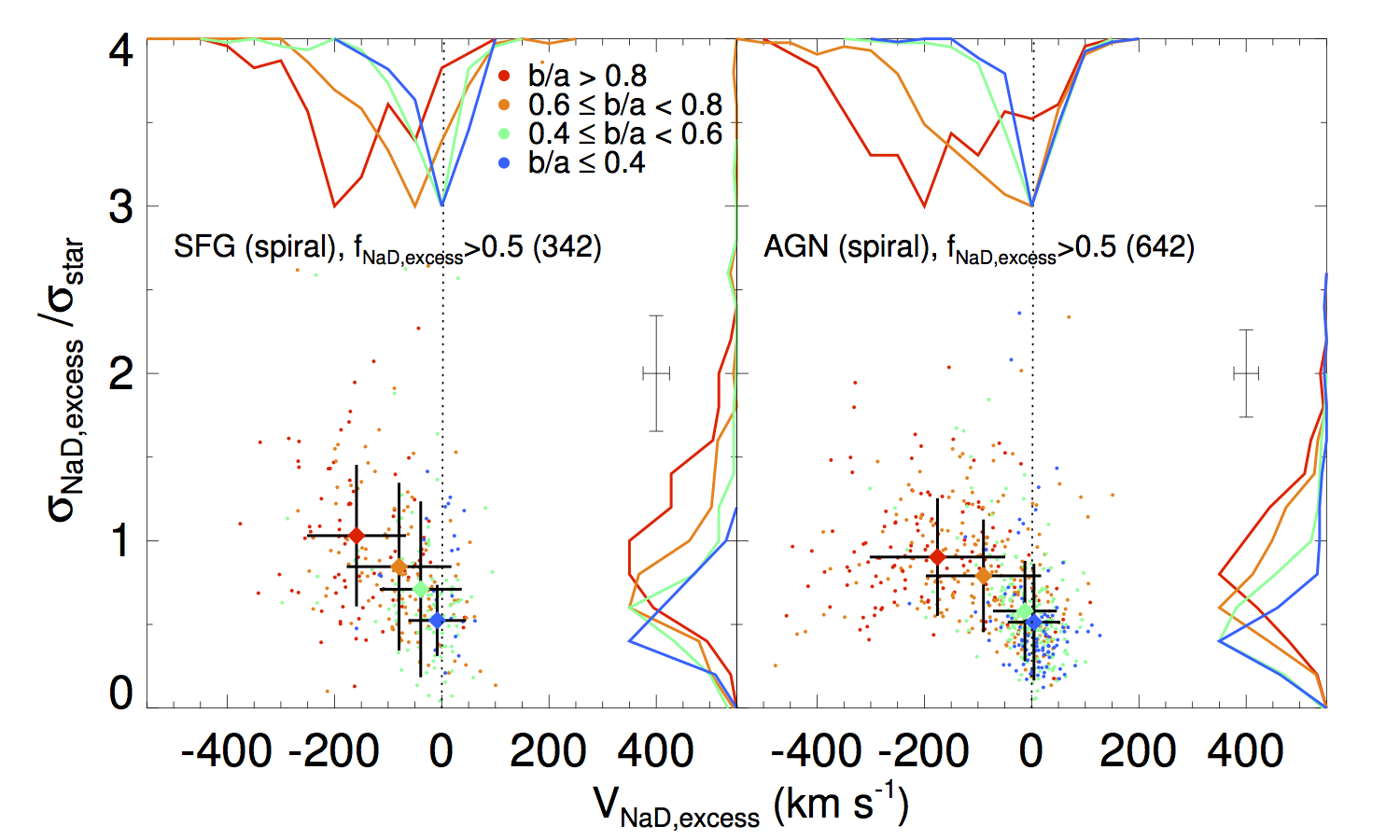}
\includegraphics[width=0.48\textwidth]{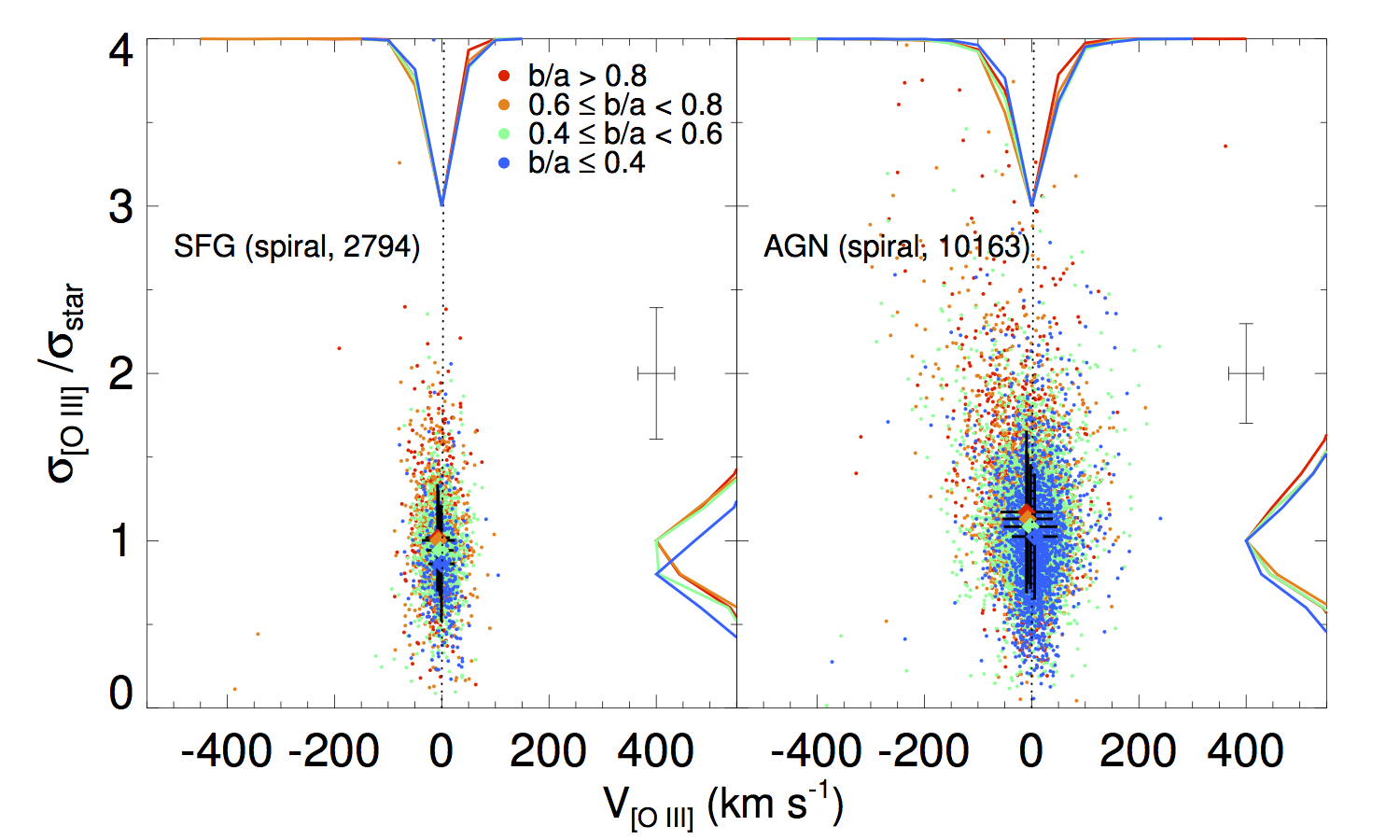}
\caption{Velocity and velocity dispersion diagram of Na D (top) and [\OIII] (bottom) for SFGs (left) and AGNs (right).
Different colors represent the minor-to-major axis (b/a) ratio: (1) b/a $>$ 0.8 (red, $\sim$face-on); (2) 0.6 $\leq$ b/a $<$ 0.8 (orange); (3) 0.4 $\leq$ b/a $<$ 0.6 (cyan); and (4) b/a $\leq$ 0.4 (blue, $\sim$edge-on). 
Normalized histograms are given in each panel. A crosshair in each panel represents a typical 1$\sigma$ uncertainty.
}
\label{fig:vvd_incl}
\end{figure}

If the outflows in the Na D excess are driven by AGNs, and the direction of outflows is random compared to host galaxy inclination. Thus, we do not expect any trend between the projected velocity of outflows and host galaxy inclination. In contrast, if the outflows are driven by SF, then the outflow direction is nearly parallel to the minor axis. In this case we expect that velocity shift and velocity dispersion of Na D excess to the line-of-sight is systematically larger for more face-on galaxies than more edge-on galaxies due to the projection effect. To test this scenario, we compare the host galaxy inclination with the velocity and velocity dispersion of the Na D excess in Figure \ref{fig:vvd_incl}. As a comparison we also investigate whether the kinematics of [\OIII] show any trend with the inclination of the host galaxies.
For this test, we only use spiral galaxies in the Na-D excess sample based on the visual classification from Galaxy Zoo project \citep{Lintott:2010gba}. These subsamples of 342 SFGs and 642 AGNs are, respectively, $\sim$28\% and $\sim$34\% of the Na-D excess sample in each group. To constrain the galaxy inclination, we use the minor-to-major axis ratio (b/a) of each galaxy. 

We find a clear dependency of the measured Na D outflow kinematics on the inclination that more face-on galaxies show on average larger blueshift and higher velocity dispersion than more edge-on galaxies (top panel in Figure 6). For example, face-on galaxies with b/a $>$ 0.8 show the average velocity shift V$_{\text{NaD,excess}}$ less than $-$150 \kms, and velocity dispersion $\sigma_{\text{NaD,excess}}$ larger than stellar velocity dispersion, while edge-on galaxies with b/a $<$ 0.4 have almost zero velocity shift and much smaller velocity dispersion, i.e., normalized $\sigma_{\text{NaD,excess}}$ by stellar velocity dispersion less than 0.5. These results clearly demonstrate that the Na D-excess kinematics measured from the line-of-sight strongly depend on the host galaxy inclination, suggesting that the outflows traced by the Na D excess are related to radial outflows (perpendicular to disk) driven by star formation, as previously reported by various studies \citep[e.g.,][]{Heckman:2000du, 2010AJ....140..445C}.

When we compare SFGs and AGNs, we find a similar trend that the strength of Na D outflows is on average higher in more face-on galaxies, indicating that the direction of the Na D outflows in AGNs is also perpendicular to the major axis of galaxies. This similar dependency may be interpreted as an evidence that the outflows manifested by Na D excess in AGN sample are also driven by the same mechanism, that derives Na D outflows in SFGs. In other words, Na D outflows detected in AGN sample is not due to AGN activity. However, it is possible to interpret the inclination dependency due to the dominance of quiescent gas in
the Na D line in edge-on galaxies. Since the Na D absorption requires a background continuum and a relatively high column density to secure the neutral state of the gas. Thus, for edge-on galaxies, Na D line will be more contaminated by quiescent (non-outflowing) gas in the disk,  \citep[for more discuss, see][]{Heckman:2000du, le11}. Thus, even if there is AGN-driven Na D outflows close to the line-of-sight, the outflow kinematics will be diluted by the contribution from quiescent gas in the disk in the case of more edge-on galaxies. Therefore, the similar inclination dependency may not be a strong evidence that Na D outflows are driven by the same mechanism in SFGs and AGNs. 

In the case of the [\OIII] kinematics we find no such inclination dependency (bottom panel in Figure 6), as expected from the case of AGN outflows since the direction of the outflows is random with respect to the major axis of the galaxies. The random orientation between AGN driven outflows and host galaxy inclination is well demonstrated by the Monte Carlo simulations by \cite{2016ApJ...828...97B}, where the measured distribution of [\OIII] velocity and velocity dispersion of a large sample of SDSS AGNs is well reconstructed using a random distribution of the angle between the outflow direction and the host galaxy major axis. Note that the difference of the inclination dependency between the kinematics of the Na D absorption line and the [\OIII] emission line is partially due to the intrinsic nature of generating Na D absorption since a background continuum is required for absorption lines, but not for emission lines.
 
\section{Conclusion}
\label{conclusion}
By utilizing a large sample of emission line galaxies selected from the SDSS at low redshift (z$<$0.3), we examined the kinematics of the Na D excess as a tracer of neutral gas outflows in the ISM. 
We found signatures of strong neutral gas outflows in the galaxies with the Na D excess, regardless of AGN activity, which is consistent with a recent study on the neutral and ionized gas kinematics for radio AGNs \citep{Sarzi:2015hf}, while the advantages of this study is that we used an unbiased sample of AGNs and SFGs, regardless of the host galaxy's properties, e.g., radio activity. The fact that the kinematic signatures manifested by Na D excess is similar between SFGs and AGNs, may suggest that neither AGN nor star-formation activities are a dominant energy source of the observed neutral gas outflows in the ISM. 
To further constrain the ambiguity, we compared the kinematics of neutral and ionized gas outflows in AGNs and SFGs. The main results indicate that the kinematics of the Na D excess and [\OIII] are independent to each another, suggesting that AGN outflows have no or little impact on the neutral gas that is traced by the Na D excess. This conclusion is supported by the recent theoretical studies that showed that AGN outflows tend to escape via a low-density channel in the ISM, e.g., ionized gas \citep{2014MNRAS.441.1615G,Bieri:2017hk}. By comparing the Na D-excess kinematics with host galaxy's inclination, we found that the primary driver of the kinematics manifested by the Na D excess is possibly star formation-driven outflows, of which the direction
is along the rotational axis of the host galaxies \citep{Heckman:2000du}.

\acknowledgments
We thank the anonymous referee for his/her valuable comments, which improved the quality of paper. The work was support by the NRF grant funded by the Korea government (No. 2016R1A2B3011457). Funding for the SDSS and SDSS-II has been provided by the Alfred P. Sloan Foundation, the Participating Institutions, the National Science Foundation, the U.S. Department of Energy, the National Aeronautics and Space Administration, the Japanese Monbukagakusho, the Max Planck Society, and the Higher Education Funding Council for England. The SDSS Web Site is http://www.sdss.org/.

\bibliographystyle{apj}

\end{document}